# The overall mortality caused by COVID-19 in the European region is highly associated with demographic composition: A spatial regression-based approach

**Srikanta Sannigrahi**[a*], **Francesco Pilla**[a], **Bidroha Basu**[a], **Arunima Sarkar Basu**[a]

[a] School of Architecture, Planning and Environmental Policy, University College Dublin Richview, Clonskeagh, Dublin, D14 E099, Ireland.

*Corresponding author: **Srikanta Sannigrahi**

E-mail:    (Srikanta Sannigrahi*)   : srikanta.sannigrahi@ucd.ie

# The overall mortality caused by COVID-19 in the European region is highly associated with demographic composition: A spatial regression-based approach

**Abstract**

The demographic factors have a substantial impact on the overall casualties caused by the COVID-19. In this study, the spatial association between the key demographic variables and COVID-19 cases and deaths were analyzed using the spatial regression models. Total 13 (for COVID-19 case factor) and 8 (for COVID-19 death factor) key variables were considered for the modelling. Total five spatial regression models such as Geographically weighted regression (GWR), Spatial Error Model (SEM), Spatial Lag Model (SLM), Spatial Error_Lag model (SEM_SLM), and Ordinary Least Square (OLS) were performed for the spatial modelling and mapping of model estimates. The local $R^2$ values, which suggesting the influences of the selected demographic variables on overall casualties caused by COVID-19, was found highest in Italy and the UK. The moderate local $R^2$ was observed for France, Belgium, Netherlands, Ireland, Denmark, Norway, Sweden, Poland, Slovakia, and Romania. The lowest local $R^2$ value for COVID-19 cases was accounted for Latvia and Lithuania. Among the 13 variables, the highest local $R^2$ was calculated for total population ($R^2 = 0.92$), followed by death crude death rate ($R^2 = 0.9$), long time illness ($R^2 = 0.84$), population with age >80 ($R^2 = 0.59$), employment ($R^2 = 0.46$), life expectancy at 65 ($R^2 = 0.34$), crude birth rate ($R^2 = 0.31$), life expectancy ($R^2 = 0.31$), Population with age 65-80 ($R^2 = 0.29$), Population with age 15-24 ($R^2 = 0.27$), Population with age 25-49 ($R^2 = 0.27$), Population with age 0-14 ($R^2 = 0.23$), and Population with age 50-65 ($R^2 = 0.23$), respectively. This suggests that the total population, death crude death rate, long time illness are the key factors that are regulating overall casualties of COVID-19. This study found that the demographic composition of the country predominantly controls the high rate of mortality and casualties due to COVID-19. In this study, the influence of the other controlling factors, such as environmental conditions, socio-ecological status, climatic extremity, etc. have not been considered. This could be scope for future research.

**1. Introduction**

The global pandemic caused by Coronavirus (COVID-19), a new genre of acute respiratory syndrome coronavirus 2 (SARS-CoV-2), has become a global health concerns for its unpredictable nature and lack of adequate medicines (WHO, 2020; Ma et al., 2020; Gorbalenya et al., 2020). Since no medicine is available yet to diagnose this novel disease, the rate of mortality and casualties due to COVID-19 is unimaginably rising worldwide from its first emergence in December 2019 in Wuhan, China. However, according to WHO, 2020, the rate of COVID-19 deaths depends on the immunity of a person, as most of the COVID-19 infected persons have experienced mild to moderate respiratory unwellness and cured without requiring special treatment. As of May 02, 2020, 3 272 202 cases and 230 104 deaths of COVID-19 reported in 215 Countries (including the applied case definitions adopted for COVID-19 and various testing strategies adopted by different countries) (WHO, 2020). Considering its surmount impact on overall human development, the United Nations, 2020, declared the disease as a social, human, and economic crisis. Most of the developing countries are experiencing the impacts and burden of this virus on the national economy. However, the negative socioeconomic consequences of COVID-19 are not only limited to developing countries, but the disease morbidity had also severely impacted the western developed countries as well (United Nations, 2020). The Congressional Research Service (2020) analyzed the economic impact of COVID-19 and predicted a 24% reduction of annual global gross domestic product (GDP), a 13% to 32% decline in global trade (Mollalo et al., 2020).

The demographic factor plays a crucial role in shaping the pattern of COVID-19 positive cases and deaths across the globe. According to UNDESA (2019) and WHO (2020), the inter(national) migrants, especially those involved in low-income jobs, are the most affected and vulnerable to death and infection of COVID-19. As of 22 April 2020, the migrants accounted for 10% of the total population for 10 out of the 15 countries having the highest number of COVID-19 cases. However, in many cases, the migrants performed a crucial role in tackling the COVID-19 emergency by working in several critical sectors (UN DESA, 2020; WHO, 2020). Ageing factor is also found crucial in controlling COVID-19 deaths and spreading. The high number of COVID-19 deaths and infection in Italy may be linked with the demographic structure of the country. The median age of the population in Italy is 46 years, and nearly a quarter of its population over the age of 65 earmarked the country as number 4$^{th}$ with a higher proportion of the old age population.  The same pattern is evident in Spain (the median age of the population is 43.9 years and more than 25000 COVID-19 deaths reported so far in Spain) (Slate, 2020; Population Europe, 2020). According to WHO (2020), in Europe, more than 95 percent of people who have died due to COVID-19 have been over 60. The long-time illness and existing respiratory disease history are also found associated with COVID-19 deaths. Zhou et al., (2020) study in Wuhan, China found that patients with existing respiratory diseases, including hypertension, diabetes, and coronary heart disease, etc. are the most vulnerable to COVID-19 deaths. Almagro & Orane-hutchinson, (2020) developed a regression model to evaluate the statistical significance between the control (neighborhood characteristics, occupations) and response variables (COVID-19 incidence) in New York City's neighborhoods. This study found that occupations were substantially explaining the observed COVID-19 patterns as people with high-level social outreach and higher social interaction were more vulnerable to be infected to the virus. Several other studies have also evaluated the association between the explanatory variables such as neighborhood characteristics (Borjas, 2020); age structure (Dowd et al., 2020 Kulu & Dorey, 2020); psychological interventions (Duan & Zhu, 2020); pre-existing health records (Fu et al., 2020); population flows and control measures (Kraemer et al., 2020); the influence of social and economic ties (Mogi & Spijker, 2020) and COVID-19 cases and deaths across the globe.

This study further advances the assessment of the impact of demographic parameters on the spread of COVID-19 cases and deaths across Europe by adopting spatial regression-based approaches. Spatial regressions models have been used extensively in many virus studies ranges from local to a global scale (Zhao et al., 2020; Guo et al., 2020; Liu et al., 2020; Liu et al., 2020). Diuk-Wasser et al., (2006) evaluated the spatial distribution of mosquito vectors for West Nile virus in Connecticut, the USA using logistic regression models. Kauhl et al., (2015) have evaluated the spatial distribution of Hepatitis C virus infections and associated determinants using Geographically Weighted Poisson Regression (GWPR) model. Kauhl et al., (2015) advocated the uses of Geographic Information Systems (GIS) and spatial epidemiological methods for providing viable screening interventions with identifying spatial hotspots/clusters as well as demographic and socio-economic determinants that have a strong association with the casualties caused by the virus. Linard et al., (2007) study on determinants of the geographic distribution of Puumala virus and Lyme borreliosis infections in Belgium found that the environmental and socio-economic factors play a crucial role in determining the spatial variation in disease risk. Mollalo et al., (2020) performed GIS-based spatial modelling to evaluate the impact of socioeconomic, behavioural environmental, topographic, and demographic factors on COVID-19 incidence in the continental United States and found that different explanatory variables including income inequality, median household income, the proportion of black females, and the proportion of nurse practitioners, etc. largely control the spatial distribution of COVID-19 cases in the USA. Malesios et al., (2020) study evaluated the spatiotemporal evolution patterns of the bluetongue virus outbreak on the island of Lesvos, Greece, and found a strong spatial autocorrelation between the spread of bluetongue virus and farms located nearby. Since COVID-19 a novel virus and no study is available so far that evaluate the close association between the demographic determinants and spread of COVID-19, this study has made an effort to address the mentioned research gap and to provide effective solutions for future preparedness for COVID-19 like situation.

## 2. Materials and methods

### 2.1 Data source and variable selection

Initially, a total of 28 demographic variables have been considered for the modelling and spatially explicit mapping of model estimates. The demographic data for the European region was collected from Eurostat[1]. The description of the variables chosen in this study is given in **Table. 1**. Using the regression models, including stepwise, forward, and backward regression models, a total of 13 (for COVID-19 case factor) and 8 (for COVID-19 death factor) variables were selected for the analysis (**Table. S1, S2**). Variables chosen for COVID-19 death modelling are – employment (Emplo), inactive population (Inc_Pop), infant mortality (Inf_Mor), people having a long-standing illness or health problem by educational attainment level (LIlln_Edu), poverty (Pov), crude birth rate (CBR), death and crude death rate (D_CDR), and total population (Tot_Pop), respectively. For COVID-19 case factor, a total of 13 variables were considered - total population (Tot_Pop), death and crude death rate (D_CDR), population by age group 0 -14 age (Pop0_14), population by age group 15 -24 age (Pop15_24), population by age group 25-49 age (Pop25_49), population by age group 50 -64 age (Pop50_64), population by age group 65 -79 age (Pop65_79), population by age group more than 80 (Pop>80), life expectancy at age 65 (LExp_65), employment (Emplo), crude birth rate (CBR), life expectancy (LExp), people having a long-standing illness or health problem by educational attainment level (LIlln_Edu), respectively. Additionally, partial least square regression (PLSR) and principal components regression (PCR) modelling was done to examine the model accuracy and identify the most important independent variables that could explain the maximum model variances. The COVID-19 cases and deaths data was retrieved from 31$^{st}$ December 2019 to 29$^{th}$ April 2020 from European Union Open data portal[2]. Few European countries (Albania, Andorra, Bosnia and Herzegovina, Czech Republic, Faroe Islands, Guernsey, Jan Mayen, Jersey, Liechtenstein, Macedonia, Monaco, Montenegro, San Marino, Serbia, and Turkey) were removed from the analysis due to data unavailability. After filtering, a total of 31 European countries were selected for spatial regression modelling and mapping.

---

[1] https://ec.europa.eu/eurostat/data/database

**2.2 Spatial regression modelling**

The spatial regression models (SRM) have been used extensively for evaluating demographic pattern analysis (Chi & Zhu, 2008), estimating land surface temperature (Jain et al., 2019; Chakraborti et al., 2018), urban air quality monitoring (Fang et al., 2015), ecosystem service valuation (Sannigrahi et al., 2020a; Sannigrahi et al., 2020b). Understanding the spatial effects such as spatial autocorrelation, spatial stationarity, and heterogeneity of a feature distribution is one of the fundamental applications of spatial regression models. In this study, total five spatial regression models include Geographically weighted regression (GWR), Spatial Error Model (SEM), Spatial Lag Model (SLM), Spatial Error_Lag model (SEM_SLM), and Ordinary Least Square (OLS) models were implemented to evaluate how the demographic factors are shaping the pattern of COVID-19 case/deaths across Europe. Among these five regression models, the global interaction between the demographic factors and COVID-19 cases/deaths were analyzed using OLS, SEM, SLM, and SEM_SLM models as these model are not impacted by spatial autocorrelation or homogeneity in the feature space. The local association between the control and response variables was calculated using the GWR model.

---

[2] https://data.europa.eu/euodp/en/data/dataset/covid-19-coronavirus-data

The GWR model is a local spatial regression model that assumes that traditional 'global' regression models such as OLS, SEM, SLM, etc. may not be effective enough do describe spatial variation of interactions, especially when spatial process varies with spatial context (Chen et al., 2018; Oshan et al., 2019). Unlike OLS, SEM, SLM spatial regression models, the GWR model depends on the assumption of spatial non-stationarity and heterogeneity in feature space and quantifies the locally varying parameter estimates (Fotheringham et al. 1996; Brundson et al., 2002). GWR calculates the location-specific interaction among the control and response variables after integrating the spatially referenced data layers (Brundson et al., 2002; Lugoi et al., 2019).

$$Y_i = \beta_o(a_j, b_j) + \sum_{i=1}^{k} \beta_i(a_j, b_j) X_i + \varepsilon_{ij} \qquad (1)$$

Where $Y_i$ is the response variable (COVID-19 case/death in this case), $\beta_o$, $\beta_i$, and $\varepsilon$ are the model parameters, $a$ and $b$ is the geographical coordinates (latitude and longitude), and $X$ is the explanatory variables (demographic variables). (Brunsdon et al., 1996) suggested that the GWR can easily compute locally varying parameter estimates, and thus found to be highly effective to produce detailed spatially explicit maps of locational variations in relationships.

The OLS is a type of global regression models that examine the (non)spatial relationships between the set of control and response variables with the fundamental assumption of homogeneity and spatial non-variability (Sun et al., 2020; Oshan et al., 2019; Mollalo et al., 2020; Ward and Gleditsch, 2018):

$$y_i = \beta_0 + x_i\beta + \varepsilon_i \qquad (2)$$

Where *i* and *yi* are the COVID-19 incidence parameters, *β₀* is the intercept, *xi* is the vector of selected demographic variables, *β* is the vector of regression coefficients, and *εi* is a random error. The fundamental function of OLS is to optimize the regression coefficients (*β*) by reducing the sum of squared prediction errors (Anselin and Arribas-Bel, 2013; Mollalo et al., 2020; Oshan et al., 2019). The usual OLS method assumes that the residual errors are homogenous and un-correlated and thereby the traditional OLS has proven to be inefficient when the errors are heterogeneous and spatially correlated and lead to a bias in regression coefficient estimation (Goodchild et al., 1993; Yang & Jin, 2010).

The SLM is based on a "spatially-lagged dependent variable" and assumes the close association between the response and control variables. Additionally, SLM also assumes dependency between the independent variables, which denotes that an independent variable could be influenced by another independent variable in the neighbourhood region (Z. Wu et al., 2020). Therefore, spatial lag function, which computes the influence of adjacent independent variables on another independent variable, can be used as a new independent variable in spatial regression modelling (Z. Wu et al., 2020). The SLM incorporates spatial dependency between the parameters into the regression model (Anselin, 2003; Ward and Gleditsch, 2018; Mollalo et al., 2020; Z. Wu et al., 2020).

$$y_i = \beta_0 + x_i\beta + \rho W_i y_i + \varepsilon_i \qquad (3)$$

where ρ is the spatial lag parameter, and $W_i$ is a vector of spatial weights. The weight matrix ($W$) of SLM indicating the neighbors at location $i$ and connects one independent variable to the explanatory variables in feature space (Anselin and Arribas-Bel, 2013; Mollalo et al., 2020)

The SEM assumes spatial dependence in the OLS residuals, which is generated from the OLS modelling error term as OLS, often ignoring the spatial dependent independent variables in the modelling (Guo et al., 2020; Z. Wu et al., 2020; Mollalo et al., 2020). Therefore, the residuals of OLS are decomposed into two spatial components- error term and a random error term (for satisfying the assumption in the modelling).

$$\begin{aligned} y_i &= \beta x_i + u_i \\ u_i &= \lambda w_i u_j + \varepsilon_i \end{aligned} \qquad (4)$$

where $u_i$ and $u_j$ are the error terms at locations $i$ and $j$, respectively, and $\lambda$ is the coefficient of spatial component errors.

The GWR model was performed using the ArcGIS Pro 2.5.0[3]. The other spatial regression models, i.e., SEM, SLM, OLS, SEM_SLM, were performed in GeoDaSpace software[4]. All the statistical analysis was performed in R studio[5] (an integrated development environment for R), Python, XLSTAT[6], and SPSS[7] software. Mapping and data visualization was done in ArcGIS Pro and R studio.

---

[3] https://www.esri.com/en-us/arcgis/products/arcgis-pro/resources
[4] https://geodacenter.github.io/GeoDaSpace/
[5] https://rstudio.com/

## 3. Results

The spatial distribution of COVID-19 cases and deaths are presented in **Fig. 1**. The highest number of cases were observed in Italy, France, Spain, and the UK. While minimum COVID-19 cases were accounted in Finland, Estonia, Latvia, Bulgaria, Greece, Slovakia, Hungary, Croatia, Slovenia, and Lithuania, respectively. Moderate levels of COVID-19 cases were detected in Poland, Romania, Austria, Denmark, Norway, Sweden (**Fig. 1**). Considering the COVID-19 deaths across Europe, the maximum COVID-19 deaths were recorded in Italy, France, Spain, and the UK, and moderate level morbidity was observed in Germany, Belgium, Netherlands, Switzerland, and Sweden. In addition, the lower COVID-19 deaths were reported in Finland, Norway, Estonia, Latvia, Lithuania, Poland, Romania, Slovakia, Hungary, Bulgaria, Greece, Slovenia, and Austria, respectively (**Fig. 1**).

The spatially varying local $R^2$ and intercept were computed using the GWR model for both COVID-19 case and death factors (**Fig. 2**). Considering local $R^2$ for the case factor, the highest association between the demographic variables and the COVID-19 case was observed in Italy and the UK. The moderate local $R^2$ was observed for France, Belgium, Netherlands, Ireland, Denmark, Norway, Sweden, Poland, Slovakia, and Romania. The lowest local $R^2$ value for COVID-19 cases was accounted for Latvia and Lithuania (**Fig. 2**). The intercept value was found in the western European region (Portugal, Spain, France, Ireland, UK, Netherlands, Belgium, Germany, and Denmark). The association between the demographic variables and COVID-19 death was also computed, and the said association was found highest in Italy, Greece, Bulgaria, Croatia, Slovenia (**Fig. 2**). Using 13 and 8 filtered demographic variables, the GWR model explained 92% and 93% model variances for COVID-19 cases and COVID-19 deaths (**Table. 2**). The adjusted $R^2$ was found higher for the COVID-19 death factor, which suggests the accuracy of the GWR model in explaining the spatial distribution of total COVID-19 deaths and its association with the demographic structure of the country. Additionally, the high local $R^2$ value derived from the GWR model for COVID-19 cases and deaths is also exhibiting the influence of population characteristics on the spread of the Corona pandemic.

---

[6] https://www.xlstat.com/en/
[7] https://www.ibm.com/analytics/spss-statistics-software

The individual influence of the 13 final demographic variables (for COVID-19 cases) is also evaluated and presented in **Fig. 3** and **Table. 3**. Among the 13 variables, the highest local $R^2$ was calculated for Tot_Pop ($R^2 = 0.92$), followed by D_CDR ($R^2 = 0.9$), LIlln_Edu ($R^2 = 0.84$), Pop>80 ($R^2 = 0.59$), Emplo ($R^2 = 0.46$), LExp_65 ($R^2 = 0.34$), CBR ($R^2 = 0.31$), LExp ($R^2 = 0.31$), Pop_65-80 ($R^2 = 0.29$), Pop_15-24 ($R^2 = 0.27$), Pop_25-49 ($R^2 = 0.27$), Pop_0-14 ($R^2 = 0.23$), and Pop_50-65 ($R^2 = 0.23$), respectively (**Table. 3**). The adjusted $R^2$ of estimated for the selected variables followed the same pattern as observed for local $R^2$. Considering the spatial association between the Tot_Pop and COVID-19 cases, the highest local $R^2$ value was observed for Italy, the UK, Slovenia, and Croatia. The moderate association between Tot_Pop and COVID-19 case was found in Spain, France, Ireland, Germany, Belgium, Netherlands, Norway, Denmark, Sweden (**Fig. 3**), whereas the lower association between Tot_Pop and cases was accounted for Estonia and Latvia (**Fig. 3**). The local $R^2$ values estimated for Pop0_14, Pop25_49, Pop50_64, Pop65_79, Pop>80, LIlln_Edu, and D_CDR, was found minimum over the western European region, and relatively higher $R^2$ values for these variables were accounted in the northern and eastern European region (**Fig. 3**).

The spatial association between the 8 demographic variables (considered for COVID-19 death) and COVID-19 deaths were analysed and presented in **Fig. 4** and **Table. 4**. Total 5 out of the 8 independent variables including Tot_Pop ($R^2$ = 0.93), D_CDR ($R^2$ = 0.93), Pov ($R^2$ = 0.63), LIlln_Edu ($R^2$ = 0.63), and Emplo ($R^2$ = 0.48) exhibited strong association with COVID-19 deaths across the Europe. The other 3 variables, i.e. Inc_Pop ($R^2$ = 0.27), CBR ($R^2$ = 0.26), and Inf_Mor ($R^2$ = 0.24) haven't produced any significant association with COVID-19 death factor. The spatially varying Local $R^2$ values of the demographic variables were found maximum in the southern and southeastern European region (Italy, Greece, Bulgaria). While a lower spatial $R^2$ value was recorded in the western region, except the Emplo variable (**Fig. 4**).

**Fig. 5** shows the linear association between 13 (for cases) and 8 (for death) demographic variables and COVID-19 spread and deaths casualties in the European countries. For the COVID-19 case factor, the coefficient of determination ($R^2$) value was recorded as 0.79, while for the death factors, the linear model has explained 64% model variances (**Fig. 5, Table. 5**). The PLS and PCR models were also executed for examining the model variances, and it has been found that the variables chosen for the COVID case factor were performed have explained the maximum model variances (**Table. 6**). For case factors, the PLS and PCR models have explained 86% and 95% model variances, and for death factors, the PLS and PCR models have explained 62% and 96% model variances (**Table. 6**).

Using the combination of demographic variables and the GWR model, the prediction of COVID-19 cases and deaths was performed and presented in **Fig. 6**. For the case factor, 93% model accuracy was observed between the actual and predicted COVID-19 cases. For the death factor, the accuracy was 97% between the predicted and actual COVID-19 death. The GWR based prediction for both case and death factors suggesting the effectivity of spatial regression models in explaining predicting the casualties caused by any epidemic/pandemic in the future time. **Fig. 7** and **Fig. 8** explained the linear association between the demographic variables and COVID-19 cases and death. For the case factor, a total 5 out of 13 variables have been strongly associated with the spread and COVID-19 cases. While for the death scenario,

Tot_Pop and D_CDR were found to be highly associated with the COVID-19 death factor. The correlation among and between the demographic variables and COVID-19 cases and deaths are presented in **Fig. 9**. Among the causative factors, high correlation were observed for Tot_Pop, D_CDR, Pop>80, and LExp, respectively. **Fig. 10** shows the similar pattern of association as Tot_Pop, D_CDR, LIlln_Edu, LExp are found to be key determinants and explained the maximum model variances.

The overall summary of the five spatial regression models are reported in **Table. 7**. Among the 13 demographic variables chosen for case factor, the average $R^2$ was observed for Tot_Pop ($R^2 = 0.82$), followed by D_CDR ($R^2 = 0.77$), Pop>80 ($R^2 = 0.3$), LIlln_Edu ($R^2 = 0.21$), L_Exp ($R^2 = 0.16$), LExp_65 ($R^2 = 0.15$), CBR ($R^2 = 0.10$), Pop25_49 ($R^2 = 0.10$), Pop15_24 ($R^2 = 0.08$), Pop50_64 ($R^2 = 0.08$), Pop65_79 ($R^2 = 0.08$), and Pop0_14 ($R^2 = 0.06$), respectively. For death factor, the highest $R^2$ value was calculated for Tot_Pop ($R^2 = 0.71$), followed by D_CDR ($R^2 = 0.63$), LIlln_Edu ($R^2 = 0.17$), Pov ($R^2 = 0.16$), Emplo ($R^2 = 0.15$), Inc_Pop ($R^2 = 0.10$), CBR ($R^2 = 0.08$), and Inf_Mor ($R^2 = 0.06$), respectively. Considering the results of all five spatial regression models, the demographic variables explained 82% model variances for COVID-19 case factor and 71% model variances for COVID-19 death factor (**Table. 7**).

## 4. Discussion

The spatial distribution of COVID-19 deaths and positive cases were mapped, and its association with key demographic variables were evaluated to understand how the demographic structure of a country can modulate the pandemic scenario caused by the novel coronavirus. The distribution of the positive COVID-19 cases and deaths were found heterogeneous across Europe. This uneven distribution could be attributed to many corresponding factors, including demography, climatic, cultural, or socio-economic differences among the countries considered in this study. For both positive COVID-19 cases and deaths, the maximum records were observed in the western European region (Spain, Italy, France, Germany, UK, Belgium, Netherlands). While, the cases and deaths were found minimum in the Eastern (Romania, Bulgaria, Greece, Estonia, Latvia, Lithuania) and Northern European region (Norway, Finland, Sweden). Similar observation was documented in Likassa et a. (2020) study where the spatial distribution of COVID-19 cases was highly associated with case-fatality rate, and the linkages between these two variables were much stronger and reached up to 8.0% for patients with the age group of 70 to 79 years and 14.8% for patients aged >80 years. Moreover, the spatial distribution of COVID-19 cases and deaths is found unpredictable for both regions and countries. Likassa et a. (2020) also stated that the high infection death rate in China, Italy, Iran, and the USA, should be linked with the spread of previous virus outbreak. According to Kraemer et al. (2020), the human mobility factor is substantially explaining the spatial distribution of COVID-19 cases in China as the growth rates become stable or negative in some areas where strong control measures were implemented and mandatorily imposed. However, the mobility factors in the other regions where the stringent regulations were not implemented, still pose severe threats by transmitting the infection in the closest neighbours (Kraemer et al., 2020).

The spatial association between the demographic variables and COVID-19 cases and deaths were found maximum in the southern, western European regions. All the 13 demographic variables considered for spatial regression modelling produced higher local $R^2$ estimates for Italy, Spain, France, Germany, UK, Greece, Bulgaria, Belgium, Netherland, Ireland. All these countries have been affected badly in terms of the total number of cases and deaths caused by COVID-19. Conversely, the weak association between demographic variables and the COVID-19 case factor was found in the eastern European countries (Estonia,

Latvia, Finland). Several factors are responsible for this uneven distribution of spatial association. This includes the age structure of the population, ratio of the elderly population, ratio of dependent population, the socio-economic structure of the society, etc. Considering the spatial association between the demographic variables and the COVID-19 death factor, the maximum values were accounted for the southern European countries, i.e., Italy, Croatia, Greece, Slovenia, France, Spain, UK, Ireland, Norway. While slightly lower estimates were observed in the eastern European regions. The intercept values calculated for the two response variables (cases and deaths) were followed the same pattern as observed for the COVID-19 death factor. The individual influences of all the 13 and 8 demographic variables on COVID-19 cases and deaths exhibited distinctive spatial association and explained substantial model variances. Among the 13 variables chosen for COVID-19 case factors, Tot_Pop, Pop>80, LIlln_Edu, and D_CDR were correlated strongly with the case factor. For the death factor, Tot_Pop, D_CDR, Pov, LIlln_Edu explained the maximum model variances.

Demographic pattern and structure of a country can substantially modulate the overall impact of a widespread pandemic and therefore may be epidemiologically informative (Jia et al., 2020); Dowd et al., 2020; Mollalo et al., 2020; Borjas, 2020; Almagro & Orane-hutchinson, 2020). In many cases, the socio-economic determinants play a crucial role in amplifying casualties due to COVID-19. Therefore, it has been suggested that paying more attention to controlling (inter)national migration, restricted population flows, modernizing the healthcare system by improving diagnosis and treatment capacity, and upgrading the public welfare system to make it fully functional for the crisis situation, could be the point of interest in order to fight against the COVID-19 like situation effectively (Su et al., 2020). The availability of sufficient SARS-CoV-2 testing centers is also found to be important for adopting control strategies and decision making for minimizing the impact of COVID-19 on the overall socio-ecological system (Rader et al., 2020). Rader et al. ( 2020) also reported the that in the USA, a total of 6,236 unique SARS-CoV-2 testing sites is available for the 3,108 counties and nearly 30%, 86%, and 5% of the total population, mountain population, and middle Atlantic population is lived in counties with median travel times over 20 min. The demographic parameters like population density, % of minority to total population, % of population with no-insurance and median income range of the population were found to be the main determinants of median travel time to testing sites. Moreover, the accessibility to SARS-Cov-2 testing sites in USA is increased with high population density and lower % of minority and uninsured people (Rader et al., 2020).

Using all the (non)spatial regression models including GWR, OLS, SLM, SER, SLM_SEM, PCR, PLS, MLR, the individual and collective effect of the demographic variables on COVID-19 cases and deaths were analyzed and reported. Among the 13 variables considered for COVID-19 case factor, Tot_Pop, D_CDR, Pop>80, LIlln_Edu, LExp, LExp_65, and Emplo factors were strongly associated with the COVID-19 cases across the European region. A similar association was observed between the demographic factors and COVID-19 in Wuhan, China (Wu et al., 2020). Wu et al., (2020) examined the association between pre-existing illness of the patients, including Acute Respiratory Distress Syndrome (ARDS) and Pneumonia and its association with COVID-19 and found that patients with exiting respiratory illness were more susceptible to COVID-19. The psychological status of the people, especially the old age people, is closely linked with the diagnostic of COVID 19 (Wang et al., 2020). Therefore, the combination of effective psychological interventions, including the lower level of psychological pressure and behavioural practices that boost mental health, can be used to improve the psychological status of vulnerable communities (Wang et al., 2020). Aging adults (>65) with long-term illness and incapable of household works were found highly vulnerable COVID-19 (Lakhani, 2020). Above the age 80, the proportion of deaths due to COVID-19 in Italy, Netherlands, Spain, and France was 50%, 58%, 59%, and 59%, respectively (Medfod & Trias-Llimos, 2020). These statistics signifying the inherent connections amongst the demographic composition and overall COVID-19 deaths and cases reported so far in the European region. Apart from the demographic factors, several climatic factors, including average temperature, minimum temperature, maximum temperature, rainfall, average humidity, wind speed, and air quality is also regulates the spread and casualties of COVID-19 (Bashir et al., 2020; Ma et al., 2020).

## 5. Conclusion

In this study, the spatial association between the demographic variables and COVID-19 cases and deaths was evaluated across the Europe. Several spatial regression models including GWR, OLS, SLM, SEM, etc. was performed for conducting the spatial regression modelling. Both COVID-19 cases and deaths were considered as dependent variables for the experiment. All the explanatory variables included in the spatial regression modelling produced high locally varying associations for Italy, Spain, France, UK. This can be attributed to the demographic composition of these countries as Italy has the second oldest population in the world and the oldest in Europe. The population composition of the other European countries i.e. Spain, France, and the Netherlands, that affected badly by COVID-19, is also dominated by senior and old age populations, thereby increasing the vulnerability to COVID-19 and many similar COVID-19 pandemics in the future (Medfod & Trias-Llimos, 2020). Lippi et al., (2020) stated that three main determinants – male sex, population with age >60, and pre-existing comorbidities such as diabetes, hypertension, chronic respiratory diseases, cancer, and cardiovascular disorders, strongly determine the rate of COVID-19 death and infection. Therefore, Lippi et al., (2020) suggested for adopting the following protection measures including "(i) minimization of direct contact with health professionals, friends, and relatives, by using digital devices, (ii) use of new technologies to intervene remotely in order to reduce the negative effects of social isolation, as well as (iii) providing timely population-specific health information to support patients and healthcare providers" to tackle the casualties of COVID-19 effectively. This study found that the high rate of mortality and casualties due to COVID-19 is predominantly controlled by the demographic composition of the country. In this study, the influence of the other controlling factors such as environmental condition, socio-ecological status, climatic extremity, etc. have not been considered. This could be a scope for future research.

**Figure Captions**

Fig. 1 The spatial distribution of COVID-19 cases and deaths and across Europe.

Fig. 2 The spatial distribution of the local $R^2$ derived from the GWR model for COVID-19 cases and death.

Fig. 3 The individual influence of the demographic variables on the COVID-19 case derived from the GWR model.

Fig. 4 The individual influence of the demographic variables on COVID-19 death derived from the GWR model.

Fig. 5 The linear association between demographic variables and COVID-19 case and death.

Fig. 6 The predicted values of COVID-19 case and death derived from the GWR model.

Fig. 7 The linear association between 13 demographic variables and the COVID-19 case.

Fig. 8 The linear association between 8 demographic variables and COVID-19 death.

Fig. 9 The correlation among and between the demographic variables and COVID-19 case and death.

Fig. 10 Sankey diagram shows the individual impact of the driving factors on COVID-19 death and case.

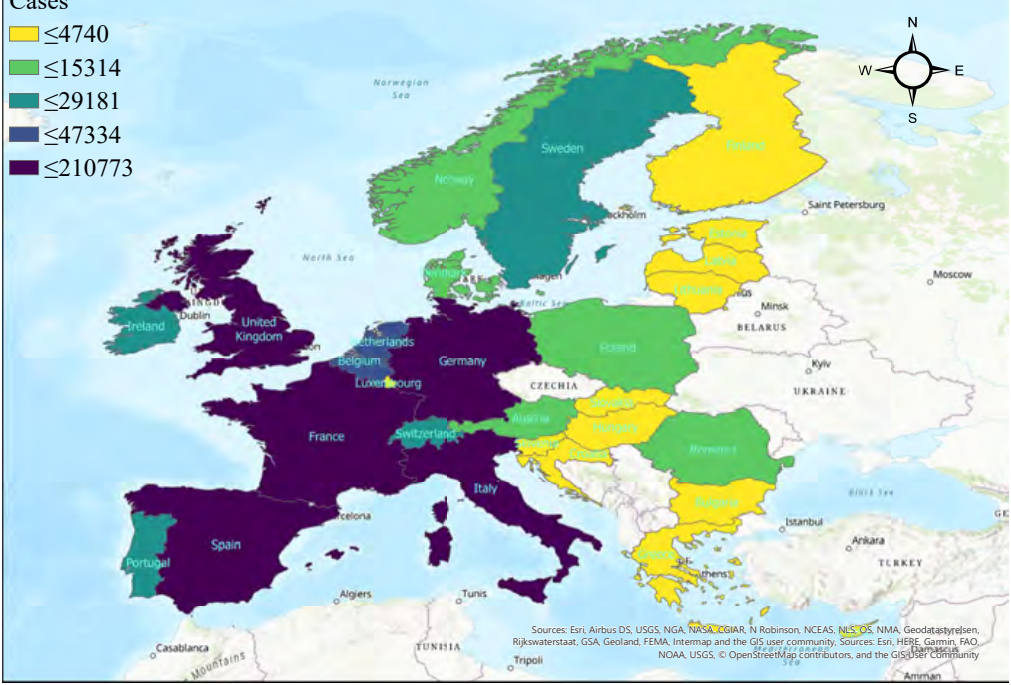
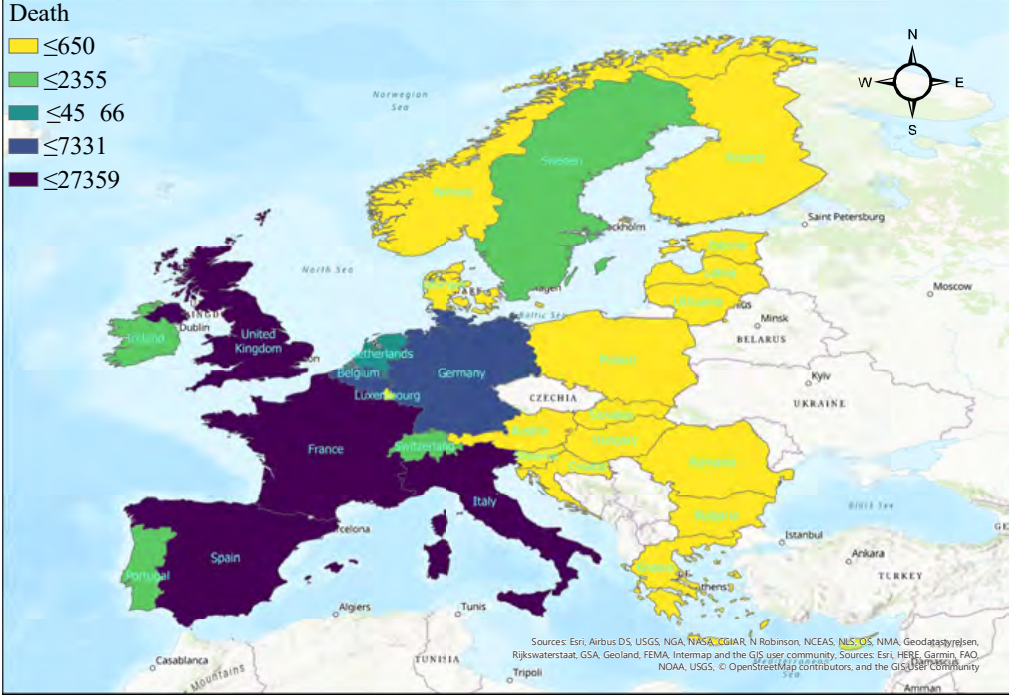

**Fig. 1**

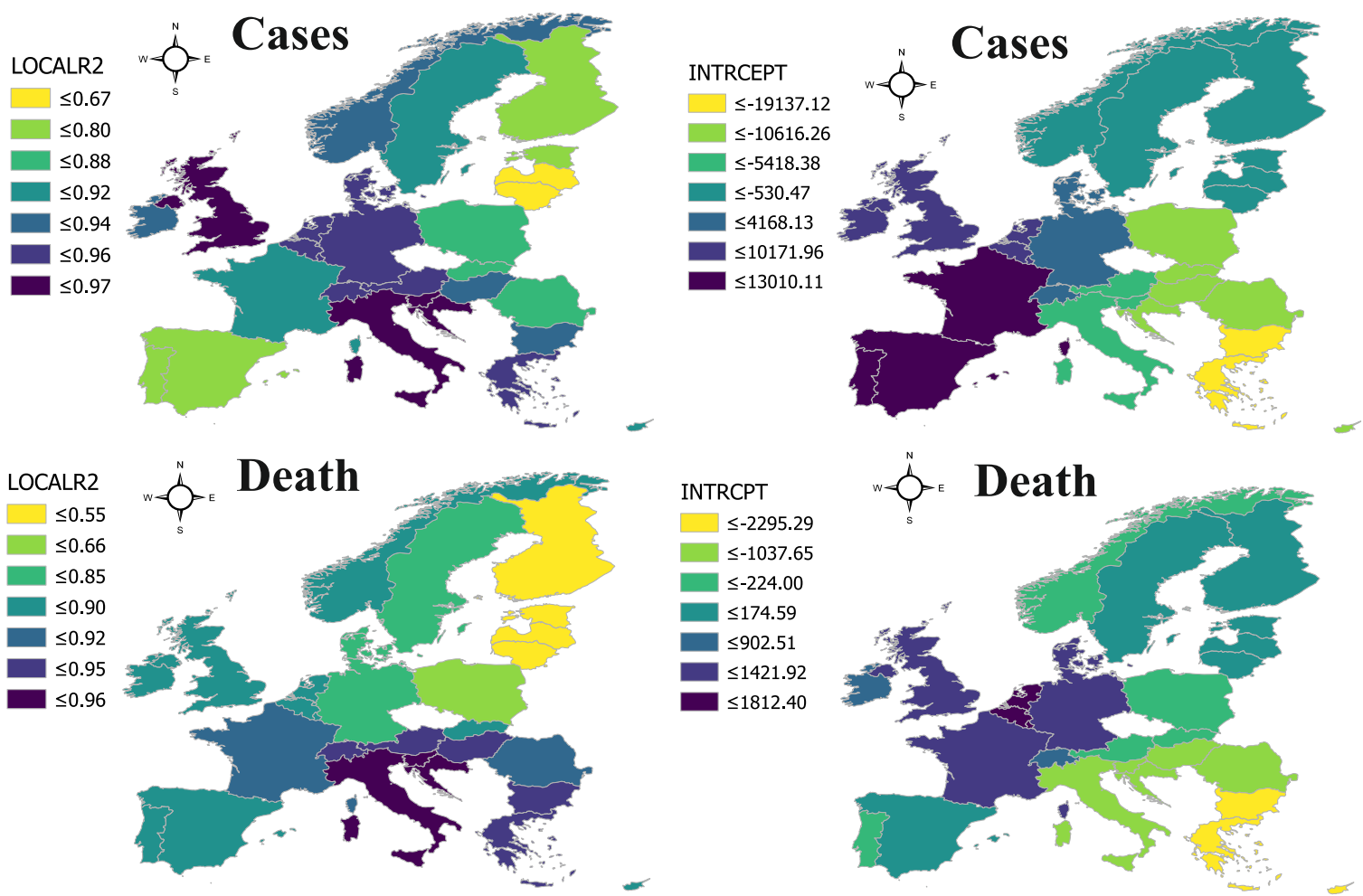

Fig. 2

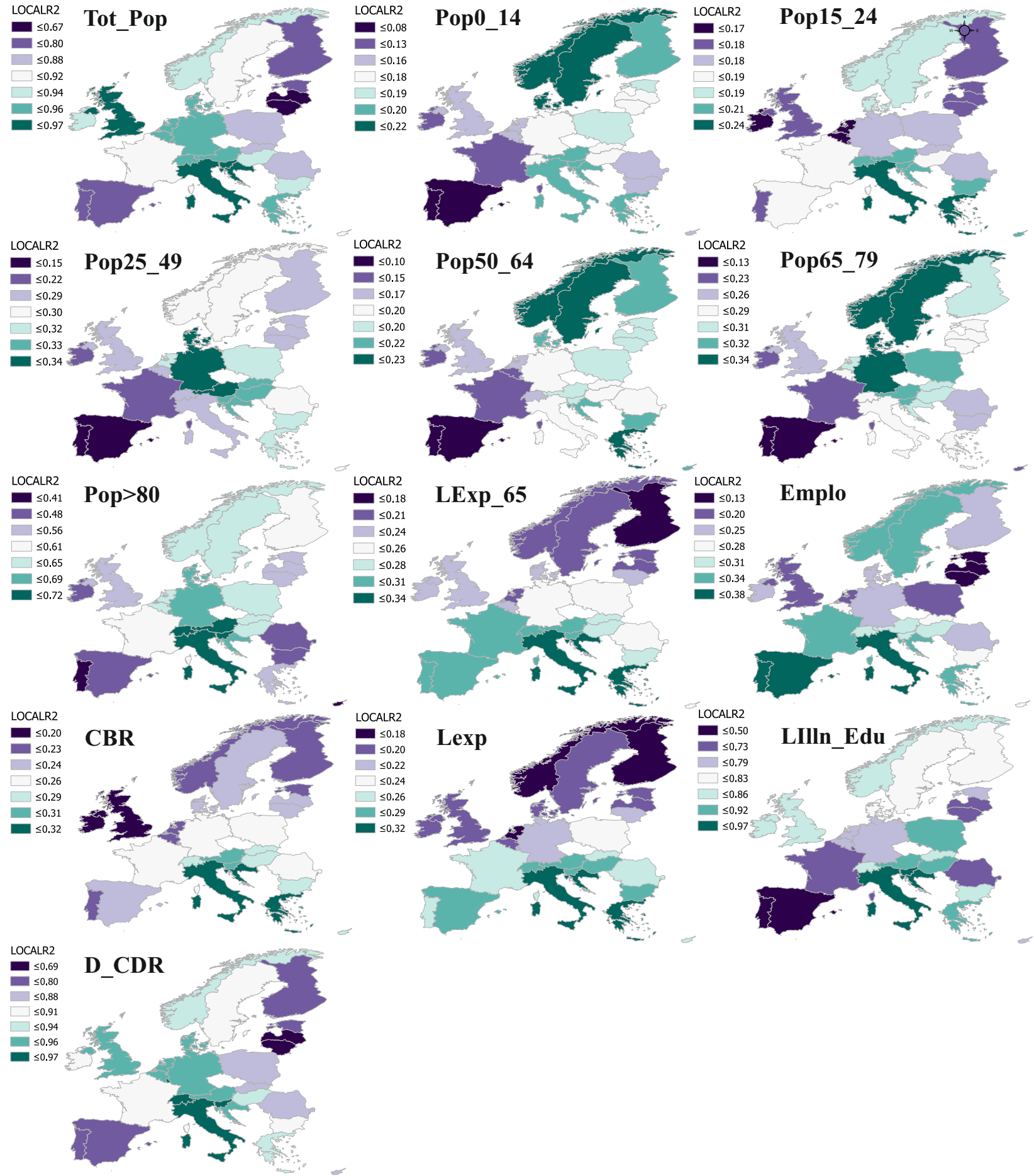

Fig. 3

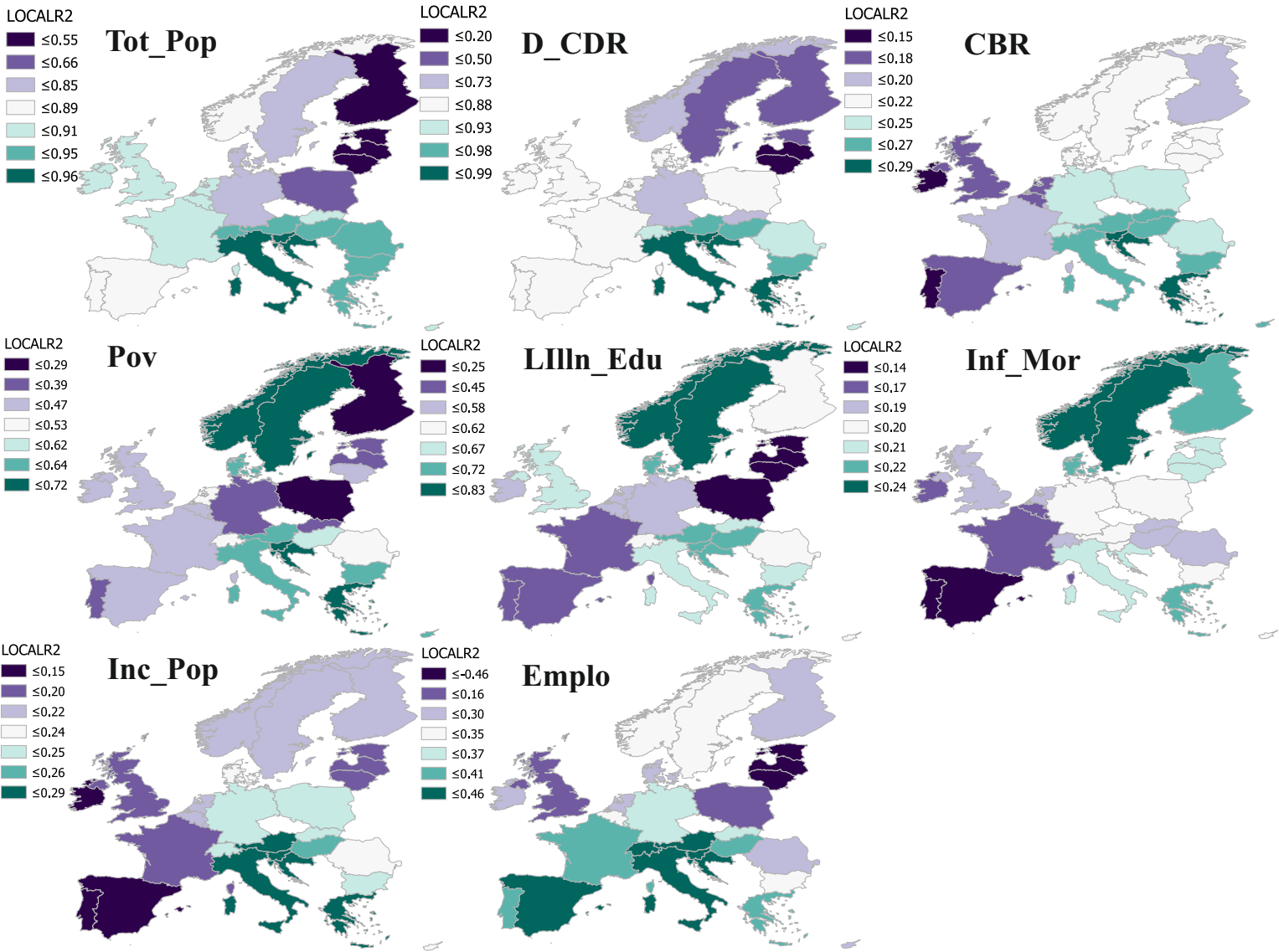

Fig. 4

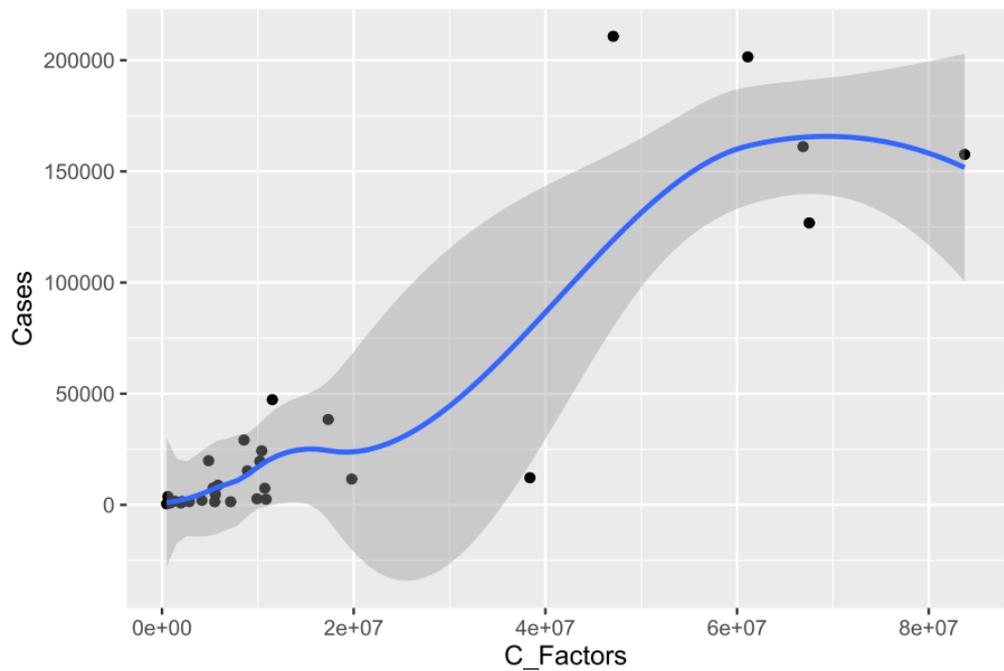
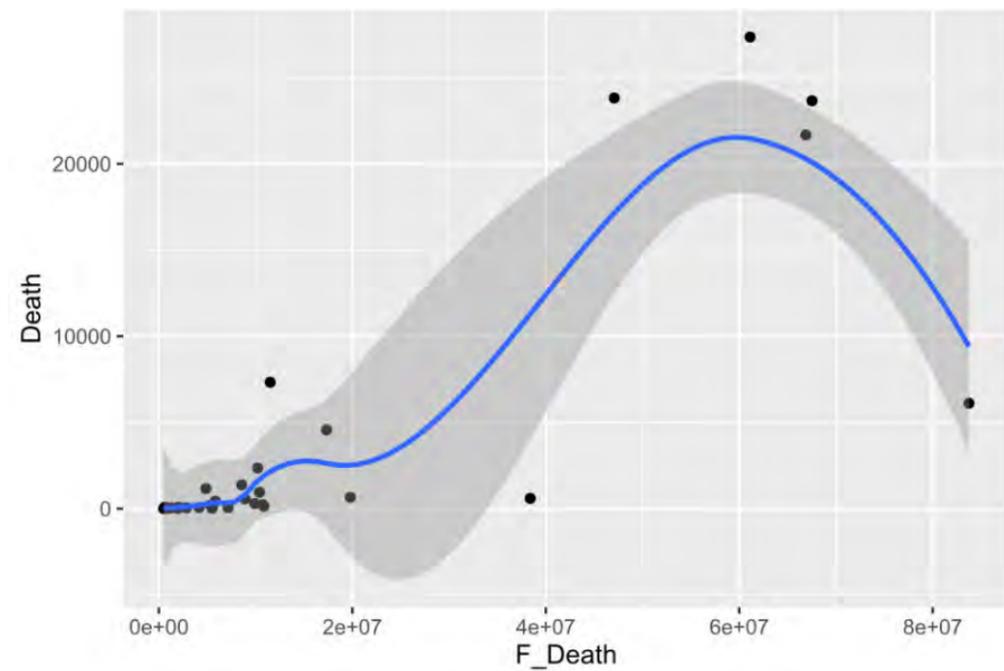

**Fig. 5**

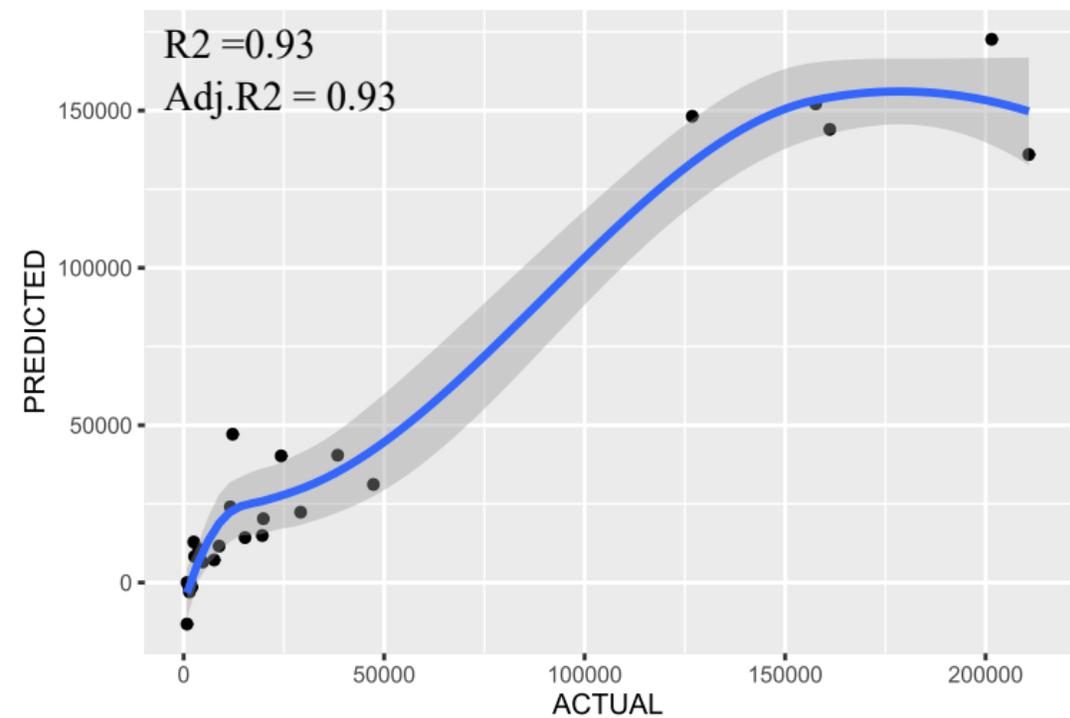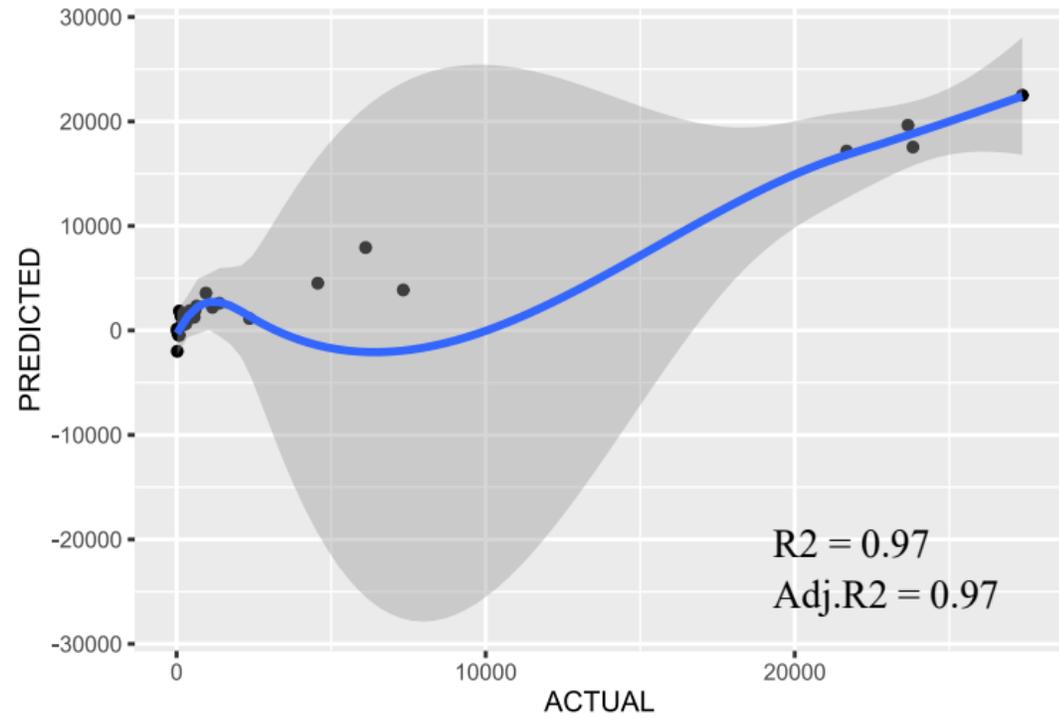

**Fig. 6**

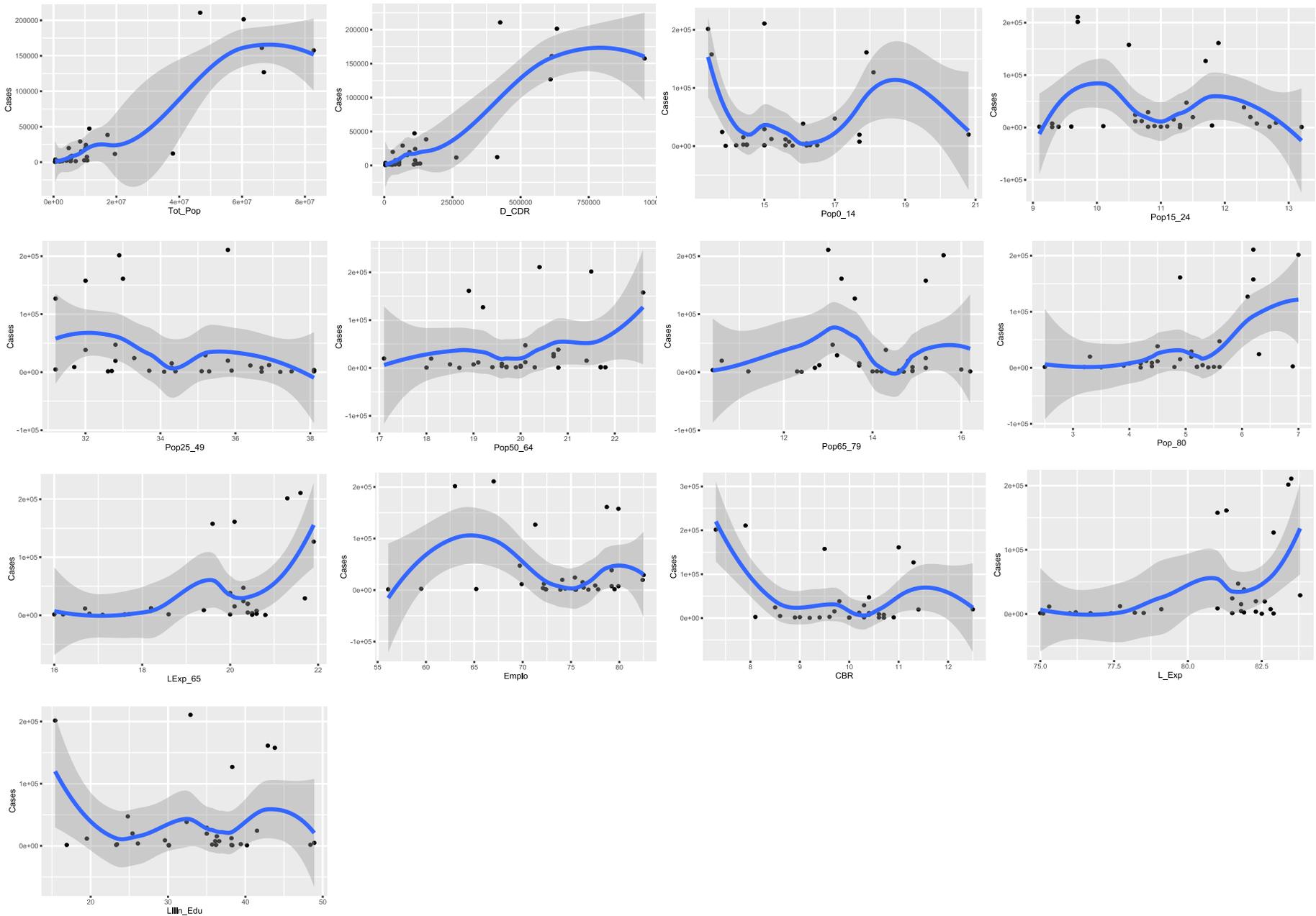

Fig. 7

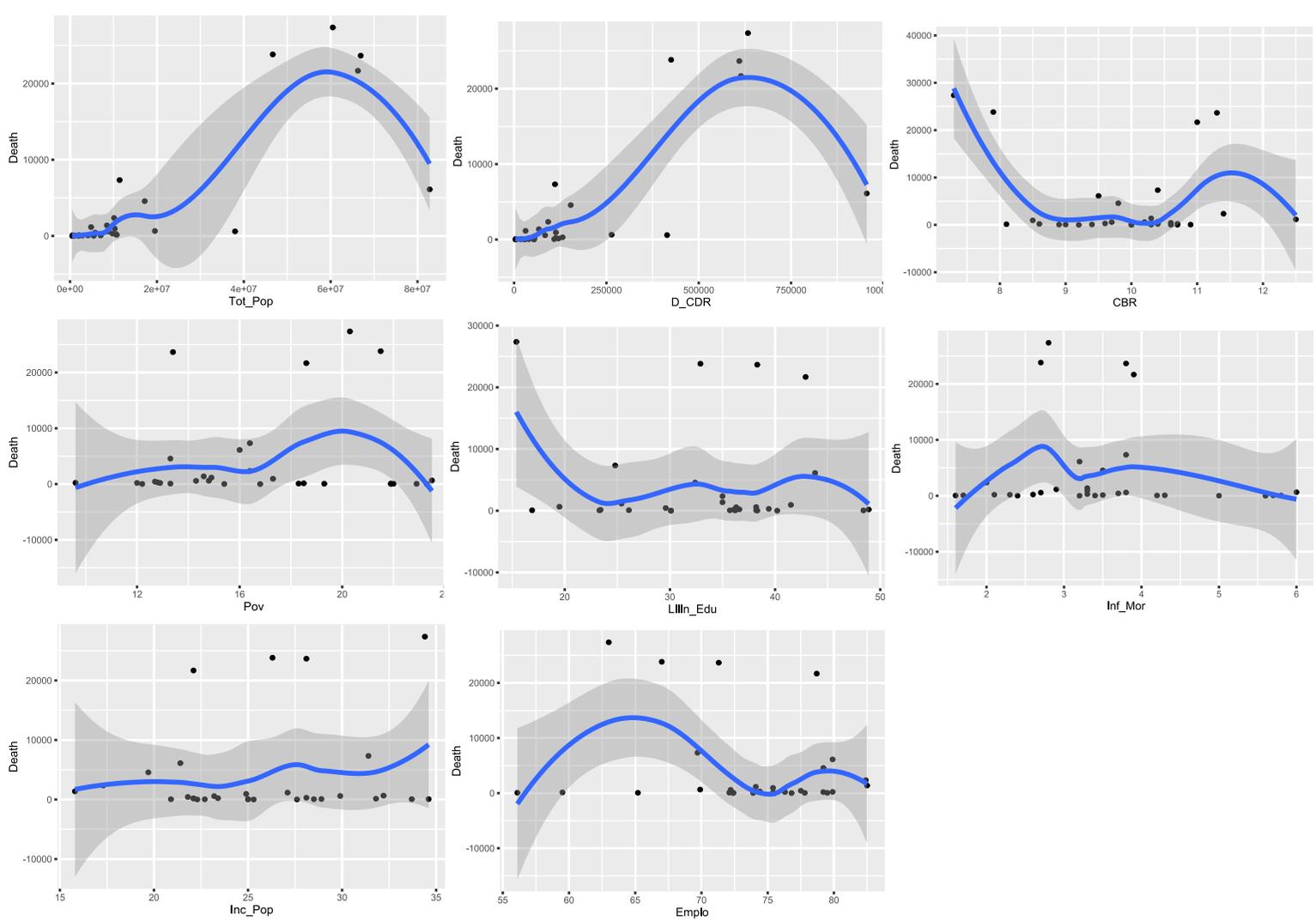

**Fig. 8**

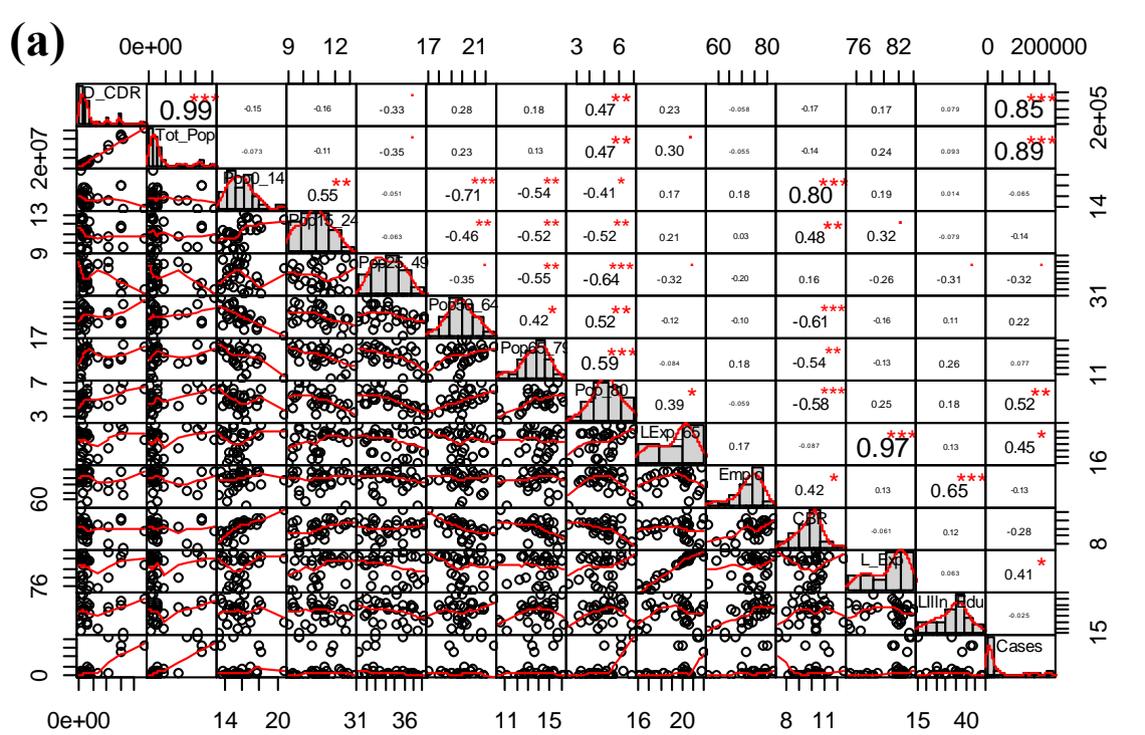

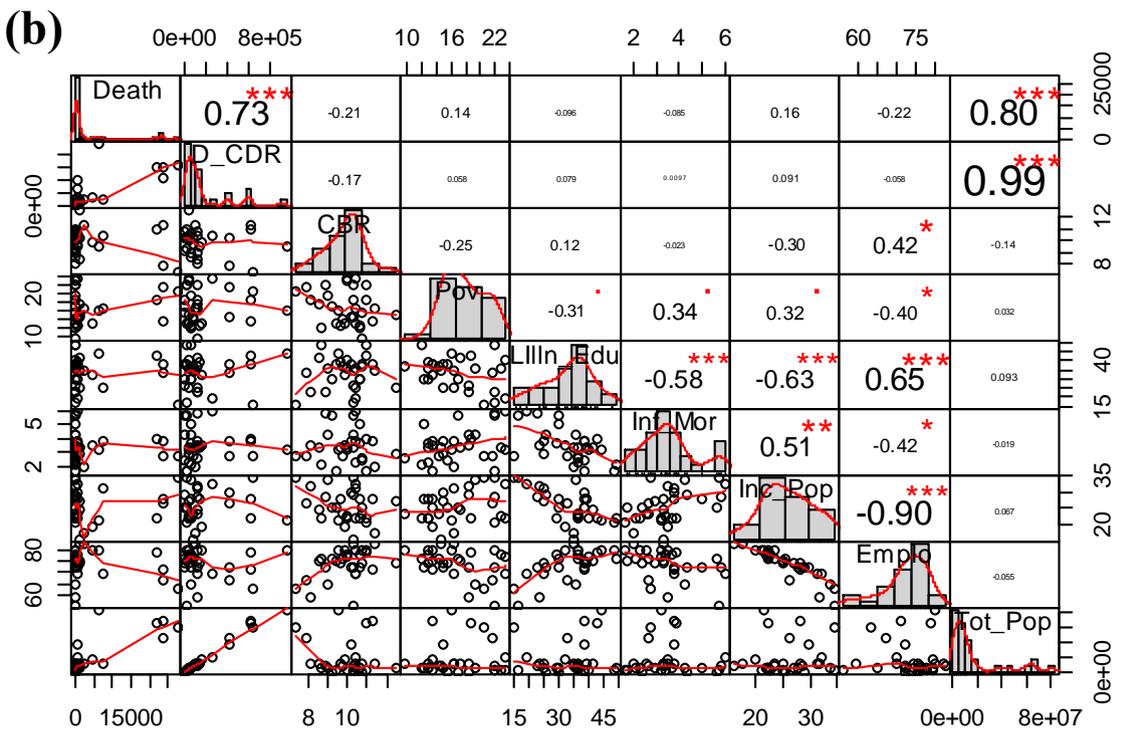

**Fig. 9**

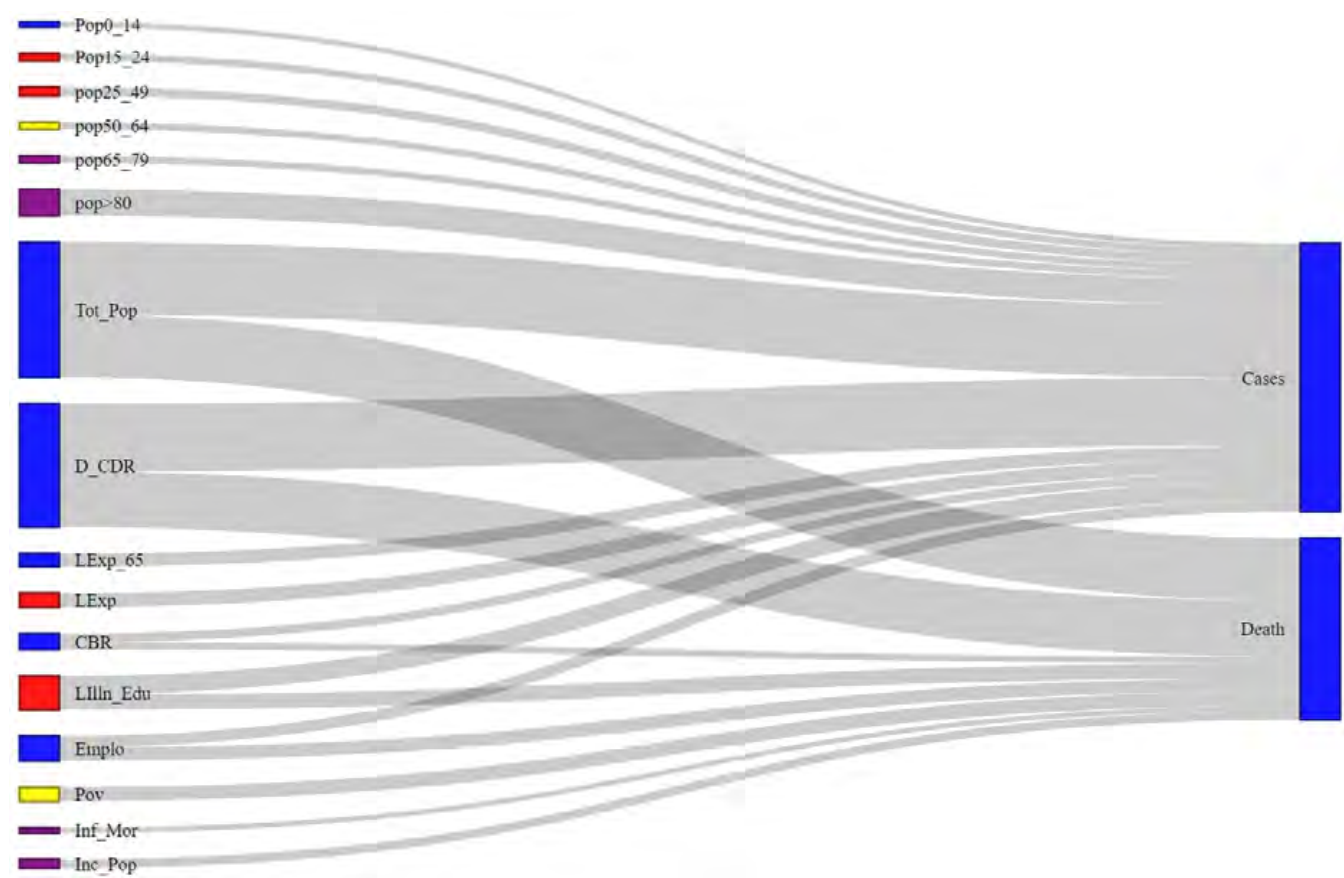

**Fig. 10**

**Table. 1** Descriptions of the demographic variables considered for the spatially explicit modelling.

| Variable Name | Code | Source |
| --- | --- | --- |
| Total Population | Tot_Pop | https://ec.europa.eu/eurostat/data/database |
| Population density | Pop_Den | https://ec.europa.eu/eurostat/data/database |
| Population by age group 0 -14 age | Pop0_14 | https://ec.europa.eu/eurostat/data/database |
| Population by age group 15 - 24 | Pop15_24 | https://ec.europa.eu/eurostat/data/database |
| Population by age group 25 - 49 | Pop25_49 | https://ec.europa.eu/eurostat/data/database |
| Population by age group 50 - 64 | Pop50_64 | https://ec.europa.eu/eurostat/data/database |
| Population by age group 65 - 79 | Pop65-79 | https://ec.europa.eu/eurostat/data/database |
| Population by age group >80 | Pop>80 | https://ec.europa.eu/eurostat/data/database |
| Life expectancy at age 65 | LExp_65 | https://ec.europa.eu/eurostat/data/database |
| Infant mortality rate | Inf_Mor | https://ec.europa.eu/eurostat/data/database |
| Proportion of population aged 65 and over | ProPop>65 | https://ec.europa.eu/eurostat/data/database |
| Deaths and crude death rate number | D_CDR | https://ec.europa.eu/eurostat/data/database |
| Old-age-dependency ratio (population 65 and over to population 15 to64 years) | OADR | https://ec.europa.eu/eurostat/data/database |
| Life expectancy at birth by sex Less than 1 year | LExp_Bir | https://ec.europa.eu/eurostat/data/database |
| Mean and median income by age and sex | Inc | https://ec.europa.eu/eurostat/data/database |
| At-risk-of-poverty rate | Pov | https://ec.europa.eu/eurostat/data/database |
| Employment rates | Emplo | https://ec.europa.eu/eurostat/data/database |
| Inactive population | Inc_Pop | https://ec.europa.eu/eurostat/data/database |
| Crude birth rate | CBR | https://ec.europa.eu/eurostat/data/database |
| Crude death rate | CDR | https://ec.europa.eu/eurostat/data/database |
| Life expectancy | L_Exp | https://ec.europa.eu/eurostat/data/database |
| People having a long-standing illness or health problem by labour status | LIlln_Lab | https://ec.europa.eu/eurostat/data/database |
| People having a long-standing illness or health problem, by educational attainment level | LIlln_Edu | https://ec.europa.eu/eurostat/data/database |
| People having a long-standing illness or health problem, by income quintile | LIlln_Inc | https://ec.europa.eu/eurostat/data/database |
| People having a long-standing illness or health problem, by degree of urbanisation | LIln_Urb | https://ec.europa.eu/eurostat/data/database |
| Self-perceived long-standing limitations in usual activities by labour status | SLoLim_Lab | https://ec.europa.eu/eurostat/data/database |
| Self-perceived long-standing limitations in usual activities due to health problem by educational attainment level | SLoLim_Edu | https://ec.europa.eu/eurostat/data/database |
| Self-perceived long-standing limitations in usual activities due to health problem by income quintile | SLoLim_Inc | https://ec.europa.eu/eurostat/data/database |

**Table. 2** Overall explanatory power of the GWR model.

| Factors | Response variable | R² | Adj.R² | Sigma² | Sigma² MLE | AICc |
|---|---|---|---|---|---|---|
| VIFCASE | CASES | 0.92 | 0.86 | 560568109.6 | 343187740.8 | 659.6173 |
| VIFDEATH | DEATH | 0.93 | 0.8758 | 8323027.718 | 5095474.638 | 541.7394 |

**Table. 3** The association between the demographic variables and total COVID-19 cases across the Europe derived from GWR model.

| Variable | Response variable | R² | Adj.R² | Sigma² | Sigma² MLE | AICc |
|---|---|---|---|---|---|---|
| TOT_POP | CASES | 0.92 | 0.8592 | 560052968.3 | 342847022.3 | 659.5954 |
| POP_0_14 | CASES | 0.23 | 0.05 | 3830253897 | 3127652440 | 704.3065 |
| POP_15_24 | CASES | 0.27 | 0.1 | 3654136070 | 2967140215 | 703.3229 |
| POP_25_49 | CASES | 0.27 | 0.11 | 3600440766 | 2980536664 | 702.3162 |
| POP_50_65 | CASES | 0.23 | 0.05 | 3847458430 | 3132947638 | 704.7236 |
| POP_65_80 | CASES | 0.29 | 0.09 | 3673442545 | 2908588079 | 703.9243 |
| POP>80 | CASES | 0.59 | 0.41 | 2371640677 | 1677393926 | 695.0507 |
| LEXP_65 | CASES | 0.34 | 0.21 | 3202906649 | 2701112406 | 698.6876 |
| EMPLOY | CASES | 0.46 | 0.24 | 3035286436 | 2218136778 | 700.6222 |
| CBR | CASES | 0.31 | 0.14 | 3485255700 | 2822255006 | 701.9217 |
| LEXP | CASES | 0.31 | 0.17 | 3336919939 | 2807628214 | 699.9052 |
| LILL_EDU | CASES | 0.84 | 0.66 | 1336243596 | 654031581.8 | 696.1295 |
| D_CDR | CASES | 0.9 | 0.84 | 644849992.9 | 398200140.7 | 663.1382 |

**Table. 4** The association between the demographic variables and COVID-19 deaths across the Europe derived from GWR model.

| Variables | Dependent variable | R² | Adj.R² | Sigma² | Sigma² MLE | AICc |
|---|---|---|---|---|---|---|
| TOT_POP | DEATH | 0.93 | 0.88 | 8291292.95 | 5075671.88 | 541.64 |
| D_CDR | DEATH | 0.93 | 0.87 | 8881154.91 | 4782300.58 | 549.63 |
| CBR | DEATH | 0.26 | 0.08 | 62658027.24 | 50738581.70 | 589.40 |
| POV | DEATH | 0.63 | 0.35 | 43358177.50 | 25422040.19 | 591.64 |
| LILL_EDU | DEATH | 0.63 | 0.41 | 39609887.61 | 25368312.11 | 583.70 |
| INF_MOR | DEATH | 0.24 | 0.08 | 62528776.91 | 51867426.87 | 588.96 |
| INC_POP | DEATH | 0.27 | 0.11 | 60695963.03 | 50238886.30 | 588.00 |
| EMPLO | DEATH | 0.48 | 0.28 | 48982608.05 | 35795674.21 | 585.08 |

**Table. 6** PCR and PLS derived model estimates showing the strong association between demographic variables and COVID-19 spread and death in Europe.

| Model | Target | Model description | R² | MSE | RMSE |
|---|---|---|---|---|---|
| PLS | Cases | cases = -105694.77+2.02*Tot_Pop+2488.22*Pop0-145210.42*Pop15_24+9.90*Pop25_49+2049.22*Pop50_64-5150.88*Pop65-79+7882.016*Pop>80+3955.45*LExp_65-1033.42*Emplo-3690.87*CBR+2205.73*L_Exp-710.87*LIlln_Edu | 0.86 | 544893412.281 | 23342.952 |
| PCR | Cases | cases=-14085739.13-0.52*D_CDR+7.67*Tot_Pop+138972.51*Pop0-14+134271.44*Pop15_24+137709.53*Pop25_49+130764.55*Pop50_64+123473.44*Pop65-79+156116.16*Pop>80-29395.17*LExp_65+4040.41*Emplo-24090.69*CBR+14241.91*L_Exp-1639.46*LIlln_Edu | 0.95 | 367054683.782 | 19158.671 |
| PLS | Death | death = 13427.22+0.01*D_CDR-718.45*CBR+136.11*Pov-42.73*LIlln_Edu-273.44*Inf_Mor+121.89*Inc_Pop-130.62*Emplo+1.32*Tot_Pop | 0.62 | 24374066.185 | 4937.010 |
| PCR | Death | death = -31265.56-0.14*D_CDR-1642.66*CBR+390.06*Pov-265.05*LIlln_Edu-1382.10*Inf_Mor+613.24*Inc_Pop+520.97*Emplo+1.68*Tot_Pop | 0.96 | 3880580.381 | 1969.919 |

**Table. 5** Details of OLS estimates for COVID-19 cases and death factors.

| Criteria | Cases | Death |
|---|---|---|
| $R^2$ | 0.7809 | 0.6283 |
| Adj. $R^2$ | 0.7725 | 0.614 |
| S.E Regression | 30999.85 | 5240.126 |
| Sigma $^2$ | 9.61E+08 | 27458917.55 |
| F | 92.6608 | 43.9484 |
| P | 4.66E-10 | 4.95E-07 |
| AICc | 660.523 | 560.975 |
| SIC | 663.187 | 563.64 |
| VIF | 2.067 | 2.067 |

**Table. 7** Overall summary of spatial regression models that indicates the linkages between the demographic variables and total COVID-19 cases and deaths across Europe.

| Target | Variables | SLM | SEM | SLM_SEM | GWR | Average |
|---|---|---|---|---|---|---|
| Cases | D_CDR | 0.73 | 0.72 | 0.73 | 0.90 | **0.77** |
|  | Tot_Pop | 0.79 | 0.78 | 0.79 | 0.92 | **0.82** |
|  | Pop0_14 | 0.01 | 0.01 | 0.01 | 0.23 | 0.06 |
|  | Pop15_24 | 0.02 | 0.02 | 0.01 | 0.27 | 0.08 |
|  | Pop25_49 | 0.02 | 0.08 | 0.02 | 0.27 | 0.10 |
|  | Pop50_64 | 0.03 | 0.04 | 0.02 | 0.23 | 0.08 |
|  | Pop65-79 | 0.00 | 0.01 | 0.00 | 0.29 | 0.08 |
|  | Pop>80 | 0.19 | 0.26 | 0.15 | 0.59 | **0.30** |
|  | LExp_65 | 0.02 | 0.22 | 0.02 | 0.34 | **0.15** |
|  | Emplo | 0.02 | 0.04 | 0.01 | 0.46 | **0.13** |
|  | CBR | 0.01 | 0.08 | 0.01 | 0.31 | 0.10 |
|  | L_Exp | 0.01 | 0.17 | ----- | 0.31 | **0.16** |
|  | LIlln_Edu | 0.01 | 0.01 | 0.01 | 0.84 | **0.21** |
| Death | Tot_Pop | 0.63 | 0.63 | 0.64 | 0.93 | **0.71** |
|  | D_CDR | 0.53 | 0.52 | 0.53 | 0.93 | **0.63** |
|  | CBR | 0.01 | 0.04 | 0.01 | 0.26 | 0.08 |
|  | Pov | 0.00 | 0.02 | 0.00 | 0.63 | **0.16** |
|  | LIlln_Edu | 0.02 | 0.02 | 0.02 | 0.63 | **0.17** |
|  | Inf_Mor | 0.00 | 0.00 | 0.00 | 0.24 | 0.06 |
|  | Inc_Pop | 0.05 | 0.04 | 0.05 | 0.27 | 0.10 |
|  | Emplo | 0.02 | 0.09 | 0.02 | 0.48 | **0.15** |
| Cases | VIF_Case | 0.79 | 0.78 | 0.79 | 0.92 | **0.82** |
| Death | VIF_Death | 0.63 | 0.63 | 0.63 | 0.93 | **0.71** |

**Table. S1** Variable selection for COVID-19 cases factors using forward and backward regression.

| Model | R | R² | Adj. R² | Std. Error of the Estimate | Change Statistics R² Change | F Change | df1 | df2 | Sig. F Change |
|---|---|---|---|---|---|---|---|---|---|
| 1 | .980a | 0.960 | 0.852 | 24121.870 | 0.960 | 8.833 | 22 | 8 | 0.002 |
| 2 | .980b | 0.960 | 0.868 | 22742.418 | 0.000 | 0.000 | 1 | 8 | 0.993 |
| 3 | .980c | 0.960 | 0.881 | 21613.311 | 0.000 | 0.032 | 1 | 9 | 0.863 |
| 4 | .980d | 0.960 | 0.890 | 20755.060 | -0.001 | 0.144 | 1 | 10 | 0.713 |
| 5 | .979e | 0.959 | 0.897 | 20069.273 | -0.001 | 0.220 | 1 | 11 | 0.648 |
| 6 | .979f | 0.959 | 0.905 | 19354.755 | 0.000 | 0.091 | 1 | 12 | 0.768 |
| 7 | .978g | 0.957 | 0.907 | 19095.902 | -0.002 | 0.628 | 1 | 13 | 0.442 |
| 8 | .977h | 0.954 | 0.908 | 18951.464 | -0.002 | 0.774 | 1 | 14 | 0.394 |
| 9 | .976i | 0.952 | 0.910 | 18817.507 | -0.002 | 0.775 | 1 | 15 | 0.393 |
| **10** | **.973j** | **0.947** | **0.906** | **19158.671** | **-0.005** | **1.622** | **1** | **16** | **0.221** |

a. Predictors: (Constant), SLoLim_Inc, Inc, D_CDR, Pop_Den, CBR, Pov, Pop25_49, LIlln_Edu, Pop15_24, Inf_Mor, Pop50_64, Inc_Pop, LExp_65, Pop0-14, Emplo, Pop>80, L_Exp, Tot_Pop, CDR, Pop65-79, OADR, LIlln_Lab

b. Predictors: (Constant), SLoLim_Inc, Inc, D_CDR, Pop_Den, CBR, Pov, Pop25_49, LIlln_Edu, Pop15_24, Inf_Mor, Pop50_64, Inc_Pop, LExp_65, Pop0-14, Emplo, Pop>80, L_Exp, Tot_Pop, Pop65-79, OADR, LIlln_Lab

c. Predictors: (Constant), SLoLim_Inc, D_CDR, Pop_Den, CBR, Pov, Pop25_49, LIlln_Edu, Pop15_24, Inf_Mor, Pop50_64, Inc_Pop, LExp_65, Pop0-14, Emplo, Pop>80, L_Exp, Tot_Pop, Pop65-79, OADR, LIlln_Lab

d. Predictors: (Constant), SLoLim_Inc, D_CDR, Pop_Den, CBR, Pov, Pop25_49, LIlln_Edu, Pop15_24, Inf_Mor, Pop50_64, Inc_Pop, LExp_65, Pop0-14, Emplo, Pop>80, L_Exp, Tot_Pop, Pop65-79, LIlln_Lab

e. Predictors: (Constant), SLoLim_Inc, D_CDR, CBR, Pov, Pop25_49, LIlln_Edu, Pop15_24, Inf_Mor, Pop50_64, Inc_Pop, LExp_65, Pop0-14, Emplo, Pop>80, L_Exp, Tot_Pop, Pop65-79, LIlln_Lab

f. Predictors: (Constant), SLoLim_Inc, D_CDR, CBR, Pov, Pop25_49, LIlln_Edu, Pop15_24, Pop50_64, Inc_Pop, LExp_65, Pop0-14, Emplo, Pop>80, L_Exp, Tot_Pop, Pop65-79, LIlln_Lab

g. Predictors: (Constant), D_CDR, CBR, Pov, Pop25_49, LIlln_Edu, Pop15_24, Pop50_64, Inc_Pop, LExp_65, Pop0-14, Emplo, Pop>80, L_Exp, Tot_Pop, Pop65-79, LIlln_Lab

h. Predictors: (Constant), D_CDR, CBR, Pov, Pop25_49, LIlln_Edu, Pop15_24, Pop50_64, Inc_Pop, LExp_65, Pop0-14, Emplo, Pop>80, L_Exp, Tot_Pop, Pop65-79

i. Predictors: (Constant), D_CDR, CBR, Pov, Pop25_49, LIlln_Edu, Pop15_24, Pop50_64, LExp_65, Pop0-14, Emplo, Pop>80, L_Exp, Tot_Pop, Pop65-79

*j. Predictors: (Constant), D_CDR, CBR, Pop25_49, LIlln_Edu, Pop15_24, Pop50_64, LExp_65, Pop0-14, Emplo, Pop>80, L_Exp, Tot_Pop, Pop65-79*

**Table. S2** Variable selection for case factors using forward and backward regression.

| Model | R | $R^2$ | Adj. $R^2$ | Std. Error of the Estimate | Change Statistics $R^2$ Change | F Change | df1 | df2 | Sig. F Change |
|---|---|---|---|---|---|---|---|---|---|
| 1 | .985[a] | 0.969 | 0.886 | 2742.586 | 0.969 | 11.550 | 22 | 8 | 0.001 |
| 2 | .985[b] | 0.969 | 0.898 | 2585.768 | 0.000 | 0.000 | 1 | 8 | 0.989 |
| 3 | .985[c] | 0.969 | 0.908 | 2454.964 | 0.000 | 0.014 | 1 | 9 | 0.909 |
| 4 | .984[d] | 0.969 | 0.916 | 2350.982 | 0.000 | 0.088 | 1 | 10 | 0.773 |
| 5 | .984[e] | 0.969 | 0.922 | 2264.371 | 0.000 | 0.132 | 1 | 11 | 0.723 |
| 6 | .984[f] | 0.969 | 0.927 | 2184.446 | 0.000 | 0.098 | 1 | 12 | 0.759 |
| 7 | .984[g] | 0.968 | 0.932 | 2112.583 | 0.000 | 0.094 | 1 | 13 | 0.764 |
| 8 | .984[h] | 0.968 | 0.936 | 2056.871 | 0.000 | 0.219 | 1 | 14 | 0.647 |
| 9 | .983[i] | 0.965 | 0.935 | 2066.895 | -0.002 | 1.156 | 1 | 15 | 0.299 |
| 10 | .981[j] | 0.962 | 0.933 | 2090.705 | -0.003 | 1.394 | 1 | 16 | 0.255 |
| 11 | .981[k] | 0.962 | 0.936 | 2049.652 | -0.001 | 0.300 | 1 | 17 | 0.591 |
| 12 | .980[l] | 0.961 | 0.939 | 2009.446 | -0.001 | 0.262 | 1 | 18 | 0.615 |
| 13 | .980[m] | 0.960 | 0.939 | 1994.592 | -0.001 | 0.705 | 1 | 19 | 0.411 |
| 14 | .979[n] | 0.959 | 0.941 | 1967.195 | -0.001 | 0.427 | 1 | 20 | 0.521 |
| **15** | **.978[o]** | **0.957** | **0.941** | **1969.919** | **-0.002** | **1.061** | **1** | **21** | **0.315** |

a. Predictors: (Constant), SLoLim_Inc, Inc, D_CDR, Pop_Den, CBR, Pov, Pop25_49, LIlln_Edu, Pop15_24, Inf_Mor, Pop50_64, Inc_Pop, LExp_65, Pop0-14, Emplo, Pop>80, L_Exp, Tot_Pop, CDR, Pop65-79, OADR, LIlln_Lab

b. Predictors: (Constant), SLoLim_Inc, Inc, D_CDR, Pop_Den, CBR, Pov, Pop25_49, LIlln_Edu, Pop15_24, Inf_Mor, Pop50_64, Inc_Pop, LExp_65, Pop0-14, Emplo, Pop>80, L_Exp, Tot_Pop, Pop65-79, OADR, LIlln_Lab

c. Predictors: (Constant), SLoLim_Inc, D_CDR, Pop_Den, CBR, Pov, Pop25_49, LIlln_Edu, Pop15_24, Inf_Mor, Pop50_64, Inc_Pop, LExp_65, Pop0-14, Emplo, Pop>80, L_Exp, Tot_Pop, Pop65-79, OADR, LIlln_Lab

d. Predictors: (Constant), SLoLim_Inc, D_CDR, CBR, Pov, Pop25_49, LIlln_Edu, Pop15_24, Inf_Mor, Pop50_64, Inc_Pop, LExp_65, Pop0-14, Emplo, Pop>80, L_Exp, Tot_Pop, Pop65-79, OADR, LIlln_Lab

e. Predictors: (Constant), SLoLim_Inc, D_CDR, CBR, Pov, Pop25_49, LIlln_Edu, Pop15_24, Inf_Mor, Pop50_64, Inc_Pop, LExp_65, Pop0-14, Emplo, Pop>80, Tot_Pop, Pop65-79, OADR, LIlln_Lab

f. Predictors: (Constant), SLoLim_Inc, D_CDR, CBR, Pov, Pop25_49, LIlln_Edu, Pop15_24, Inf_Mor, Pop50_64, Inc_Pop, LExp_65, Pop0-14, Emplo, Pop>80, Tot_Pop, Pop65-79, OADR

g. Predictors: (Constant), SLoLim_Inc, D_CDR, CBR, Pov, Pop25_49, LIlln_Edu, Pop15_24, Inf_Mor, Pop50_64, Inc_Pop, Pop0-14, Emplo, Pop>80, Tot_Pop, Pop65-79, OADR

h. Predictors: (Constant), SLoLim_Inc, D_CDR, CBR, Pov, Pop25_49, LIlln_Edu, Pop15_24, Inf_Mor, Pop50_64, Inc_Pop, Pop0-14, Emplo, Pop>80, Tot_Pop, Pop65-79

i. Predictors: (Constant), D_CDR, CBR, Pov, Pop25_49, LIlln_Edu, Pop15_24, Inf_Mor, Pop50_64, Inc_Pop, Pop0-14, Emplo, Pop>80, Tot_Pop, Pop65-79

j. Predictors: (Constant), D_CDR, CBR, Pov, Pop25_49, LIlln_Edu, Pop15_24, Inf_Mor, Pop50_64, Inc_Pop, Pop0-14, Emplo, Tot_Pop, Pop65-79

k. Predictors: (Constant), D_CDR, CBR, Pov, Pop25_49, LIlln_Edu, Pop15_24, Inf_Mor, Pop50_64, Inc_Pop, Emplo, Tot_Pop, Pop65-79

l. Predictors: (Constant), D_CDR, CBR, Pov, Pop25_49, LIlln_Edu, Pop15_24, Inf_Mor, Inc_Pop, Emplo, Tot_Pop, Pop65-79

m. Predictors: (Constant), D_CDR, CBR, Pov, Pop25_49, LIlln_Edu, Inf_Mor, Inc_Pop, Emplo, Tot_Pop, Pop65-79

n. Predictors: (Constant), D_CDR, CBR, Pov, LIlln_Edu, Inf_Mor, Inc_Pop, Emplo, Tot_Pop, Pop65-79

**o. Predictors: (Constant), D_CDR, CBR, Pov, LIlln_Edu, Inf_Mor, Inc_Pop, Emplo, Tot_Pop**